\documentclass{article}

\usepackage[final]{neurips_2025}

\usepackage[utf8]{inputenc} % allow utf-8 input
\usepackage[T1]{fontenc}    % use 8-bit T1 fonts
\usepackage{hyperref}       % hyperlinks
\usepackage{url}            % simple URL typesetting
\usepackage{booktabs}       % professional-quality tables
\usepackage{amsfonts}       % blackboard math symbols
\usepackage{nicefrac}       % compact symbols for 1/2, etc.
\usepackage{microtype}      % microtypography
\usepackage{xcolor}         % colors
\usepackage{listings}
\usepackage{multirow}

\usepackage{graphicx}

\title{Towards Efficient Multi-Scale Deformable Attention on NPU}

\author{%
Chenghuan Huang$^{1}$ \quad Zhigeng Xu$^{1}$ \quad Chong Sun$^{2}$ \quad Chen Li$^{2}$ \\
\quad \textbf{Ziyang Ma}$^1$\thanks{Corresponding author.} \\
$^1$WeChat HPC, Tencent Inc. \quad $^2$WeChat Vision, Tencent Inc. \\
\texttt{entityless@outlook.com, xuzgcs@foxmail.com, maziyang08@gmail.com} \\
\texttt{\{waynecsun,chaselli\}@tencent.com}
}

\begin{document}

\maketitle

\begin{abstract}
Multi-scale deformable attention (MSDA) is a flexible and powerful feature extraction mechanism for visual tasks, but its random-access grid sampling strategy poses significant optimization challenges, especially on domain-specific accelerators such as NPUs. In this work, we present a co-design approach that systematically rethinks memory access and computation strategies for MSDA on the Ascend NPU architecture. With this co-design approach, our implementation supports both efficient forward and backward computation, is fully adapted for training workloads, and incorporates a suite of hardware-aware optimizations. Extensive experiments show that our solution achieves up to $5.9\times$ (forward), $8.9\times$ (backward), and $7.3\times$ (end-to-end training) speedup over the grid sample-based baseline, and $1.9\times$, $2.4\times$, and $2.0\times$ acceleration over the latest vendor library, respectively.
\end{abstract}


\section{Introduction}

Multi-scale deformable attention (MSDA)~\cite{deformabledetr} has become a cornerstone in modern vision transformers, enabling efficient and flexible feature aggregation across multiple spatial resolutions. Unlike traditional attention mechanisms with quadratic complexity, MSDA leverages sparse, learnable sampling of key points from multi-scale feature maps, significantly reducing computational cost while maintaining high detection accuracy. This approach has been widely adopted in state-of-the-art object detection~\cite{deformabledetr,defa,mdha,ueda}, 3D perception~\cite{bevformer,voxformer,aibench,davinci}, and other vision tasks, driving advances in both 2D and 3D domains.

\begin{figure}[ht]
    \centering
    \includegraphics[width=0.7\linewidth]{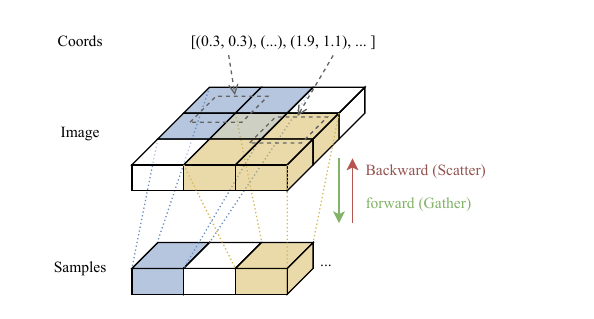}
    \caption{Grid sampling irregular memory access pattern.}
    \label{fig:grid_sample}
\end{figure}

The core strength of MSDA lies in its grid-sampling-based random access strategy, which allows each query to attend to a small, task-adaptive set of spatial locations across different feature levels. This flexibility enables models such as Deformable DETR~\cite{deformabledetr} and BEVFormer~\cite{bevformer} to achieve superior performance, especially on small or occluded objects and in multi-view scenarios. However, this same flexibility introduces significant challenges for efficient implementation, particularly on general-purpose hardware like CPUs and GPUs. The irregular memory access patterns and large memory footprint inherent in grid sampling lead to poor hardware utilization and become a major bottleneck in end-to-end inference latency~\cite{defa,ueda}; see Figure~\ref{fig:grid_sample}.

To address these challenges, recent research has explored both algorithmic and architectural optimizations. For example, DEFA~\cite{defa} proposes a co-design approach that combines pruning-assisted grid sampling with dedicated hardware acceleration, achieving substantial speedup and energy efficiency gains over conventional platforms. Other works have introduced universal deformable attention accelerators~\cite{ueda}, efficient benchmarking tools~\cite{aibench}, and performance modeling frameworks~\cite{modeling,squeeze} to further advance the deployment of MSDA in real-world applications.

Despite these efforts, deploying high-performance MSDA on domain-specific accelerators such as Ascend NPUs remains challenging. The lack of massive thread-level parallelism (TLP) and limited support for random memory access on these architectures exacerbate the inefficiencies of conventional approaches. As a result, MSDA often becomes a critical performance bottleneck in large-scale vision models, motivating the need for further research into co-design strategies that bridge the gap between algorithmic flexibility and hardware efficiency.

% \begin{figure}[ht]
%     \centering
%     \includegraphics[width=0.7\linewidth]{figures/co_design.pdf}
%     \caption{Common co-design paradigm.}
%     \label{fig:co_design}
% \end{figure}

In this paper, we propose xMSDA, a high-performance implementation of MSDA on Ascend NPUs with a co-design paradigm. Our main contributions are as follows:
\begin{itemize}
    \item \textbf{Co-design analysis and methodology:} We provide a systematic co-design analysis of MSDA and the Ascend NPU architecture, identifying key computational and memory bottlenecks through detailed profiling and microbenchmarking. This analysis guides the development of hardware-aware optimization strategies for both forward and backward passes.
    \item \textbf{Efficient implementation:} Notably, we exploit non-officially supported type-unaligned gather instructions to enable efficient and correct grid sampling, overcoming hardware documentation limitations. In addition, we introduce adaptive vectorization strategies, padding-based alignment fixes, and contention-aware scheduling for scatter operations.
    \item \textbf{Significant acceleration:} Extensive experiments demonstrate that our implementation achieves up to $5.9\times$ (forward), $8.9\times$ (backward), and $7.3\times$ (end-to-end training) speedup over the PyTorch grid-sample baseline, and up to $2.4\times$ over the latest official CANN operator. Ablation studies further validate the effectiveness of each proposed optimization.
\end{itemize}

\section{Motivation}

Although multi-scale deformable attention (MSDA) offers substantial improvements in flexibility and accuracy, its practical deployment in large-scale vision models faces significant efficiency challenges. Prior studies have shown that MSDA-related operators, such as MSDA, can dominate the computational cost in modern object detection and 3D perception networks~\cite{defa,ueda}. For example, in Deformable DETR, MSDA accounts for over 50\% of the end-to-end inference latency, and similar trends are observed in other state-of-the-art models.

In our own production-scale models, profiling results indicate that MSDA operations consume approximately 15\% of the total training time, making it one of the most time-consuming components in the training pipeline. This high overhead is primarily attributed to the grid-sampling mechanism, where each query dynamically generates sampling locations across multi-scale feature maps. The resulting random memory access patterns severely limit hardware utilization, especially on platforms lacking efficient support for irregular data movement.

From a systems perspective, the main bottleneck of MSDA lies in the unpredictable and data-dependent access patterns induced by the sampling locations. Unlike standard convolution or dense attention, where memory accesses can be statically optimized, the coordinates in MSDA are dynamically determined and change throughout training. This makes it infeasible to apply pre-computed access pattern optimizations, such as those used in DEFA~\cite{defa}, which rely on static analysis or pruning of sampling candidates.

Consequently, there is a pressing need for a general and efficient implementation of MSDA that does not rely on prior knowledge of sampling patterns. Such a solution must support both forward and backward passes with high performance, regardless of the underlying hardware support for thread-level parallelism or random memory access. Addressing this challenge is critical for unlocking the full potential of MSDA in real-world, large-scale vision applications.

\section{Co-design Analysis}

\begin{figure}[ht]
    \centering
    \includegraphics[width=0.75\linewidth]{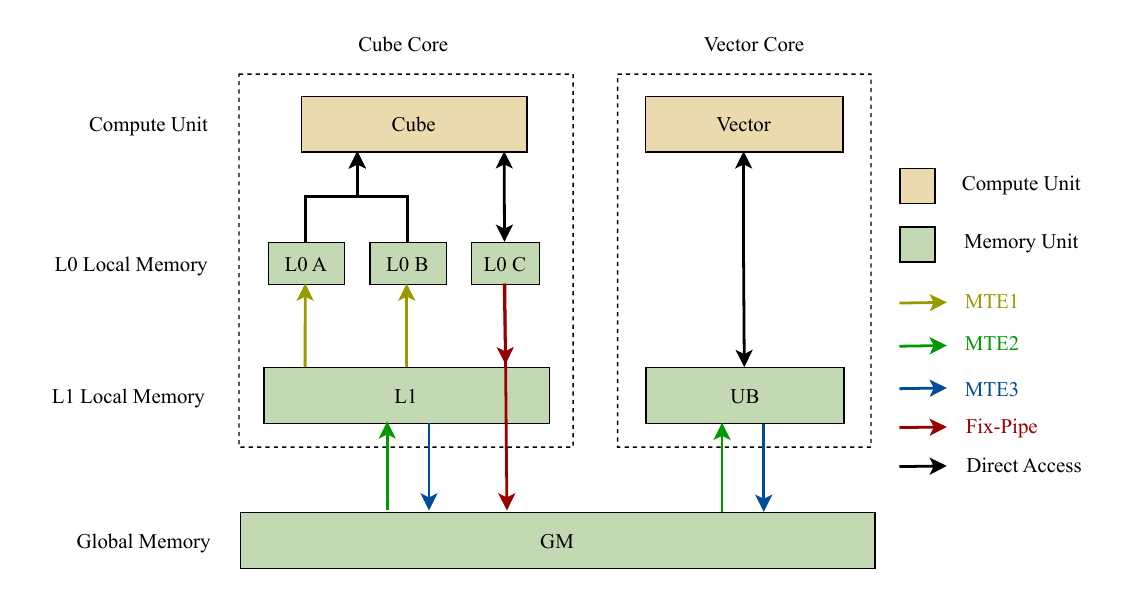}
    \caption{Ascend disaggregated AI core architecture.}
    \label{fig:ascend_npu_arch}
\end{figure}

The Ascend NPU utilizes a highly heterogeneous architecture where scalar units handle general logic, cube cores are tailored for high-throughput matrix multiplications, and vector cores are well-suited for SIMD (Single Instruction Multiple Data) operations, as shown in Figure~\ref{fig:ascend_npu_arch}. The scalar unit is not explicitly marked, as it resides in each cube and vector core. Notably, the latest Ascend NPUs, such as the 910B2C, incorporate a disaggregated cube-vector core design, resulting in non-trivial communication through global memory (GM) between cube cores and vector cores. To maximize computational intensity, each cube core and vector core adopts a hierarchical on-chip memory structure. The cube compute unit accesses basic tiled matrix-multiplication operands directly from L0 local memory (L0A, L0B, and L0C), allowing larger data block reuse in the secondary L1 local memory. In contrast, the vector compute unit directly accesses its on-chip L1 local memory, the Unified Buffer (UB), without intermediate L0 local memory. Data movement is accomplished via various Memory Transfer Engines (MTE) and Fix-Pipe, achieving asynchronous parallelism with computations.

Detailed performance specifications of the Ascend 910B2C NPU are listed in Table~\ref{tab:ascend_spec}. Since the cube core is dedicated to matrix multiplication, we mainly leverage vector cores to accelerate MSDA operations. Specifically, we further explore the capabilities of random memory accesses on vector cores. Results show that only gathering data within UB is natively supported, while gathering or scattering data between UB and GM can only be accomplished by multiple explicit fine-grained data copy operations. However, it is important to note that overlapping memory access latencies through massive thread-level parallelism is infeasible, as the Ascend NPU does not adopt an SIMT (Single Instruction Multiple Thread) design but relies on a limited number of high-performance cube and vector cores. This necessitates top-down random memory access optimizations for compute patterns such as MSDA.

\begin{table}[ht]
    \centering
    \caption{Ascend 910B2C Performance Specifications.}
    \label{tab:ascend_spec}
    \begin{tabular}{lcccc}
        \toprule
        \#Cube & \#Vec & Peak Performance (Cube@FP16) & Peak Performance (Vector@FP16) \\
        \midrule
        24 & 48 & 353 & 20 \\
        \bottomrule
    \end{tabular}
    \begin{tabular}{lccc}
        \toprule
        GM Size (GB) & UB Size (KB) & GM Bandwidth (GB/s)\\
        \midrule
        64 & 192 & 1800\\
        \bottomrule
    \end{tabular}    
\end{table}

\begin{figure}[ht]
    \centering
    \includegraphics[width=0.99\linewidth]{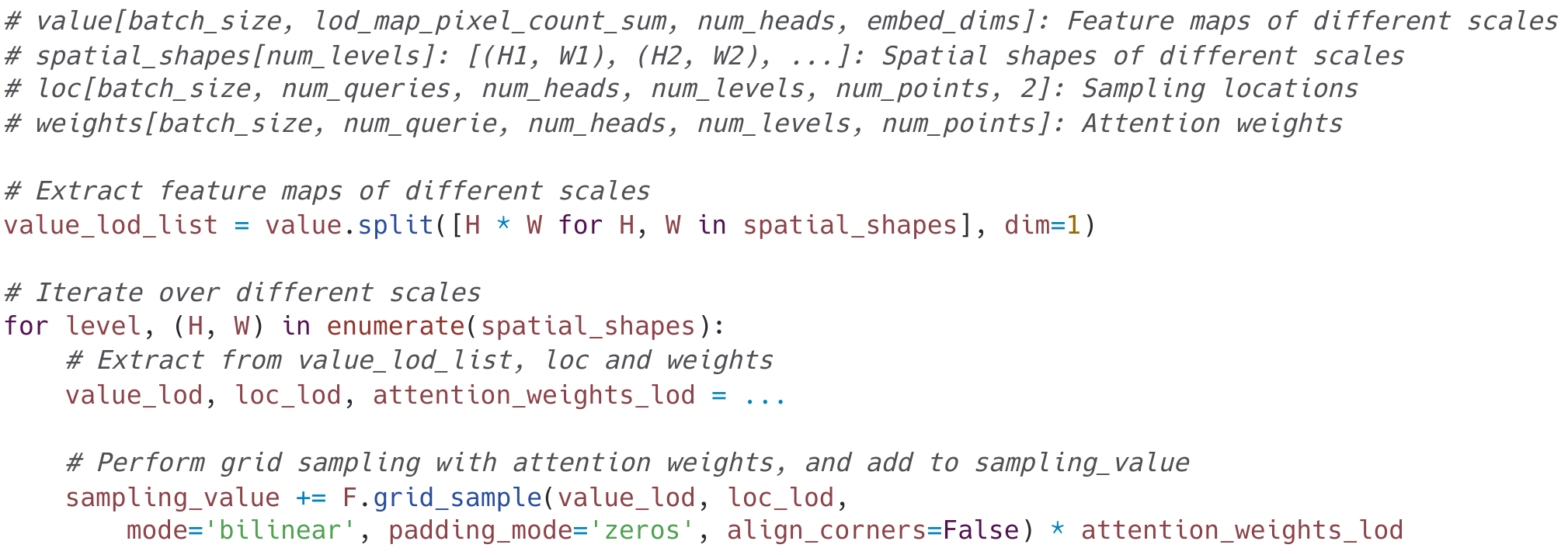}
    \caption{MSDA PyTorch implementation}
    \label{fig:msda_code_block}
\end{figure}

To better illustrate the engineering challenges and optimizations of MSDA, we first present the reference implementation from MMCV, which is widely used in the community, in Figure~\ref{fig:msda_code_block}.

The bottleneck of MSDA lies predominantly in grid sampling operations, both in the forward and backward passes. In the forward pass, the primary hotspot is a gather operation on the input feature maps, while in the backward pass, the main bottleneck is a scatter operation on the gradient values.

To ground our analysis, we first define the typical MSDA input used in our training scenarios. Specifically, we follow previous works~\cite{liu2021swin,li2024visual} and extract five feature maps from a Swin Transformer backbone, with downsampling rates of $1/4$, $1/8$, $1/16$, $1/32$, and $1/64$. Each input image is resized to $1024\times1024$, resulting in the largest feature map of size $256\times256$. All feature maps are projected to a dimension of 256. We adopt a multi-head setting with 8 heads and use 4 sampling points per query, consistent with common practice. The resulting essential input size for MSDA is $87296 \times 32$. Input data is stored in FP16 format, while all internal MSDA computations are performed in FP32 to ensure high precision.

Given the MSDA input and the Ascend NPU platform characteristics, we observe that the UB can accommodate a single-channel feature map, since the largest $256 \times 256$ feature map occupies 131,072 bytes. Therefore, in the forward pass, feature map access can be organized either as per-channel UB gather or as the simpler per-pixel GM gather across all channels. The optimal strategy can be determined in advance using microbenchmarking on the Ascend NPU platform.

For the backward scatter operation, performance profiling is also necessary to identify the optimal pattern for higher efficiency.

Before microbenchmarking, we revisit the characteristics of grid sampling. Here, we use bilinear interpolation, meaning each sampling point accesses a $2 \times 2$ spatial neighborhood. In standard row-major memory layouts, $x_0$ and $x_1$ are stored contiguously. Thus, the gather and scatter granularity on GM should be 64 elements, while the UB gather granularity is 2 elements.

We generate random offsets for read/write operations. For GM reads, the alignment granularity is $32 \times 2$ bytes; for GM writes, it is $32 \times 4$ bytes, matching the feature map's read/write alignment. We find that, except for very small GM arrays, the GM read/write performance is largely independent of array size, so we report only a representative set of results.

\begin{figure}[ht]
    \centering
    \includegraphics[width=0.99\linewidth]{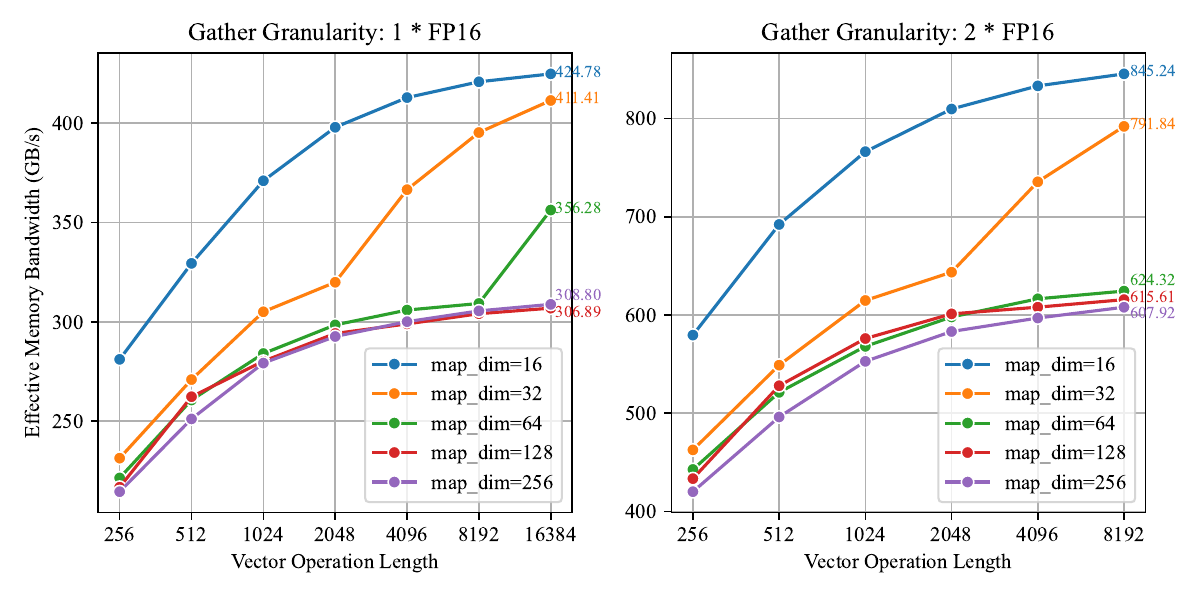}
    \caption{Microbenchmark results for UB gather with varying granularity and feature map dimensions. Each subplot corresponds to a different gather granularity.}
    \label{fig:ub_gather}
\end{figure}

Figure~\ref{fig:ub_gather} shows the microbenchmark results for UB gather operations with varying granularity and feature map dimensions. Each subplot corresponds to a different gather granularity.

As shown in Figure~\ref{fig:ub_gather}, the effective throughput for random-index UB gather reaches up to 845 GB/s, while GM gather achieves only 106 GB/s. Thus, UB gather offers a substantial advantage over GM gather. Furthermore, merging adjacent pixels for gather yields significant bandwidth improvements: before merging, the maximum UB gather bandwidth is only 424 GB/s. The gather bandwidth decreases as feature map size increases, dropping to 607 GB/s for $256 \times 256$ feature maps. Additionally, increasing the vector length (vec\_len) markedly improves bandwidth, indicating a strong positive correlation between vector length and execution efficiency on the Vector Core.

\begin{figure}[ht]
    \centering
    \includegraphics[width=0.7\linewidth]{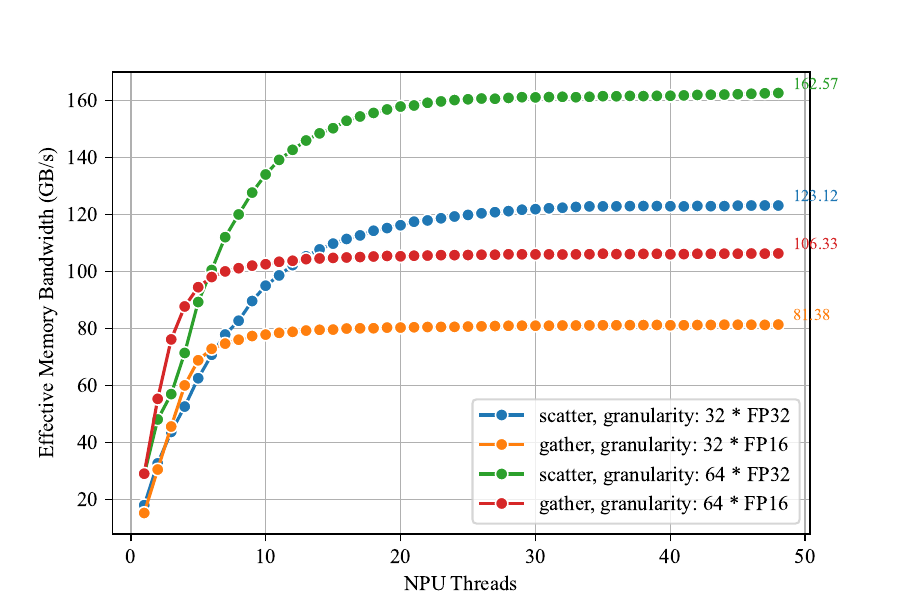}
    \caption{Microbenchmark results for GM gather and scatter with different granularities.}
    \label{fig:gm_scatter}
\end{figure}

We next examine the GM gather and scatter microbenchmark results, as shown in Figure~\ref{fig:gm_scatter}. Here, we test both gather and scatter operations with different granularities. Increasing the write granularity significantly boosts bandwidth: without merging pixels, the write bandwidth is only 106 GB/s, but after merging, it reaches 162 GB/s. Although contention is less pronounced than in GM gather, increasing the number of NPU threads still introduces some contention effects.

\section{Implementation}
\label{sec:Implementation}

The implementation of MSDA computes the sampling coordinates and attention weights, then performs a bilinear grid sampling operation for each feature level, followed by a weighted sum to produce the output. While this approach is flexible and expressive, it poses significant challenges for efficient deployment on domain-specific hardware such as Ascend NPUs, especially due to the irregular memory access patterns and the need for high-throughput gather and scatter operations.

We introduce a series of targeted optimizations for both the forward and backward passes, as detailed below.

\subsection{Forward}

The forward computation involves calculating the grid sampling weights and coordinates, followed by gathering the corresponding values from the input feature maps. In our implementation, we rearrange the layout of the value tensor so that the pixel dimension is the last axis, which can be efficiently handled using PyTorch tensor operations.

\begin{figure}[ht]
    \centering
    \includegraphics[width=0.99\linewidth]{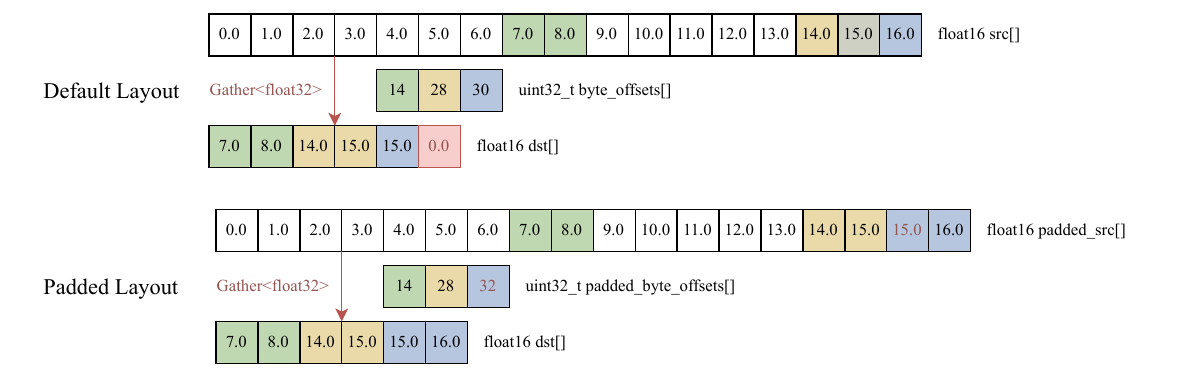}
    \caption{Error cases and fixes for type-unaligned gather behavior at specific byte offsets.}
    \label{fig:gather_error}
\end{figure}

A key challenge arises when aggregating two consecutive pixels ($x_0$, $x_1$) using a gather operation with a granularity of 4 bytes but an alignment of 2 bytes, utilizing the gather instruction for FP32 to process FP16 data. According to the official Ascend C API documentation, this is an undefined behavior. To address this, we empirically analyze the behavior of the gather operation under all possible byte offsets. As shown in Figure~\ref{fig:gather_error}, we observe that errors occur specifically when $\mathrm{idx} \% 32 = 30$, while other cases function as expected. Our solution is to pad the input feature map and correspondingly adapt the 1D indices for padding. This ensures correct gather behavior across all cases.

\begin{figure}[ht]
    \centering
    \includegraphics[width=0.7\linewidth]{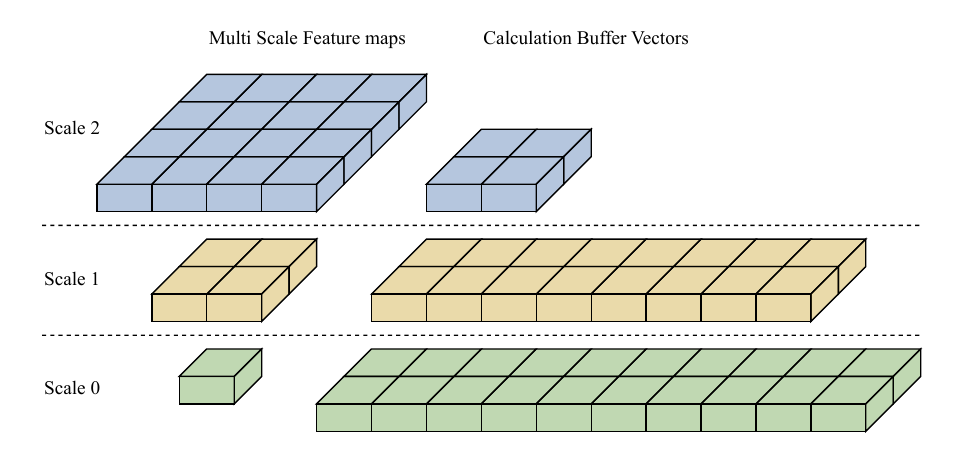}
    \caption{Adaptive vector length selection for different feature map sizes.}
    \label{fig:adaptive_vec}
\end{figure}

Additionally, since the feature maps at different levels of detail (lod) have varying sizes, the available UB for element-wise vector computation also varies. We therefore adaptively adjust the vector computation unit length for each lod to fully utilize the UB and maximize vector core efficiency, as shown in Figure~\ref{fig:adaptive_vec}.

For training scenarios, we store the gather result of feature maps for backward computation, which introduces additional IO for the forward pass.

\subsection{Backward}

The backward pass consists of two main parts: (1) computing the gradients with respect to the sampling locations and attention weights, which is a combination of standard element-wise vector operations; and (2) computing the gradient with respect to the input value, which is dominated by scatter-add operations.

At this stage, the grad value tensor is laid out with the embedding dimension as the last axis, so merging $x_0$ and $x_1$ for write-out is fully supported by the hardware architecture.

\begin{figure}[ht]
    \centering
    \includegraphics[width=0.9\linewidth]{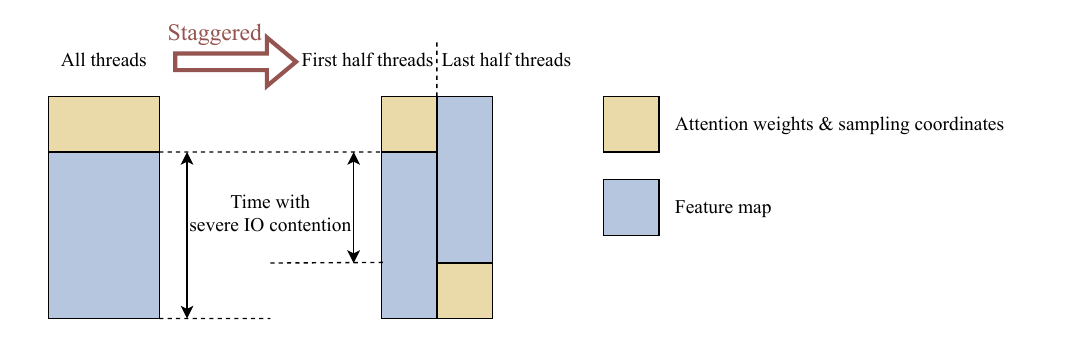}
    \caption{Staggered gradient computation phases to reduce scatter contention.}
    \label{fig:reduce_contention}
\end{figure}

Based on our earlier microbenchmark results, we know that scatter-add operations can suffer from severe contention. To mitigate this, we implement a feature that staggers the computation to reduce conflicts, as illustrated in Figure~\ref{fig:reduce_contention}. By offsetting the execution of the two core phases, we lower the overall parallelism of grad value computation, thereby reducing contention and improving effective bandwidth.

\section{Evaluation}

In this section, we comprehensively evaluate the performance of our optimized MSDA implementation on the Ascend NPU platform, comparing it against both the widely adopted PyTorch grid-sample-based baseline as DEFA~\cite{defa}, and the kernel library from the vendor's full-stack software framework called CANN (Compute Architecture for Neural Networks). We further conduct ablation studies to quantify the impact of each optimization.

\subsection{Experimental Setup}

All experiments are conducted on a server equipped with dual Intel(R) Xeon(R) Platinum 8476C CPUs and 2TB DDR4 memory, running Ubuntu 22.04.4 LTS with Linux kernel 5.4.241. The software stack includes CANN 8.1.RC1.beta1 (released 2025/04/30). We use representative MSDA input tensors from object detection training tasks, and for the backward pass, random tensors are used as gradients. All measurements are performed on the NPU, using profiler-reported kernel time (in microseconds), averaged over 100 runs after warm-up. For multi-operator processes, total kernel time is reported.

\subsection{Overall Performance}

Table~\ref{tab:overall_perf} summarizes the kernel time for forward and backward passes across different implementations. Our optimized MSDA achieves significant speedups over both the PyTorch baseline and the official CANN operator, especially in the backward pass.

% Forward table

\begin{table}[ht]
    \centering
    \caption{Forward \& backward pass kernel time (\textmu s).}
    \label{tab:overall_perf}
    \begin{tabular}{lcccc}
        \toprule
        Forward & Baseline & CANN & Ours (Inference) & Ours (Train) \\
        \midrule
        Time & 52662.7 & 16573.6 & \textbf{8981.6} & 15562.5 \\
        \bottomrule
    \end{tabular}
    \begin{tabular}{lccc}
        \toprule
        Backward & Baseline & CANN & Ours \\
        \midrule
        Time & 335696.8 & 91056.4 & \textbf{37714.1} \\
        \bottomrule
    \end{tabular}
\end{table}

% Backward table

% Speedup table

\begin{table}[ht]
    \centering
    \caption{Relative speedup over baseline and CANN.}
    \label{tab:speedup}
    \begin{tabular}{lcc}
        \toprule
        & vs. Baseline & vs. CANN \\
        \midrule
        Inference & $5.86\times$ & $1.85\times$ \\
        Backward & $8.90\times$ & $2.41\times$ \\
        Train (Fwd+Bwd) & $7.29\times$ & $\textbf{2.02}\times$ \\
        \bottomrule
    \end{tabular}
\end{table}

Our approach delivers up to $8.9\times$ speedup in the backward pass and $7.3\times$ in end-to-end training compared to the baseline, and more than $2\times$ over the official CANN operator.

\subsection{Ablation Study}

To further analyze the effectiveness of each optimization described in Section~\ref{sec:Implementation}, we conduct ablation experiments by selectively disabling them. We report both kernel time and the utilization ratio of key hardware units in Table~\ref{tab:ablation_fwd_inf} as measured by the profiler:
\begin{itemize}
    \item \textbf{Vector Ratio}: Proportion of time spent on vector computation units.
    \item \textbf{Scalar Ratio}: Proportion of time spent on scalar computation units.
    \item \textbf{MTE2}: Utilization of the memory transfer engine for GM $\rightarrow$ UB transfers.
    \item \textbf{MTE3}: Utilization of the memory transfer engine for UB $\rightarrow$ GM transfers.
\end{itemize}

% Forward inference ablation

\begin{table}[ht]
    \centering
    \caption{Ablation test results.}
    \label{tab:ablation_fwd_inf}
    \begin{tabular}{lcccc}
        \toprule
        Forward (Inference) & Default & $-$Adaptive VecLen & $-$Gather Fusion & $-$All \\
        \midrule
        Time (\textmu s) & 8981.6 & 10846.3 & 10550.4 & 16558.6 \\
        Vector Ratio (\%) & 72.56 & 67.14 & 87.92 & 67.6 \\
        Scalar Ratio (\%) & 46.12 & 77.28 & 67.2 & 96.4 \\
        MTE2 (\%) & 63.07 & 62.61 & 36.58 & 28.22 \\
        MTE3 (\%) & 36.3 & 48.63 & 20.83 & 25.9 \\
        \bottomrule
    \end{tabular}

    \begin{tabular}{lcccc}
        \toprule
        Forward (Train) & Default & $-$Adaptive VecLen & $-$Gather Fusion & $-$All \\
        \midrule
        Time (\textmu s) & 15562.5 & 18011.3 & 16555.4 & 25615.8 \\
        Vector Ratio (\%) & 40.31 & 40.11 & 53.01 & 42.6 \\
        Scalar Ratio (\%) & 41.66 & 73.03 & 67.2 & 97.2 \\
        MTE2 (\%) & 79.76 & 77.25 & 61.35 & 24.1 \\
        MTE3 (\%) & 73.72 & 70.67 & 49.98 & 31.22 \\
        \bottomrule
    \end{tabular}

    \begin{tabular}{lcccc}
        \toprule
        Backward & Default & $-$Staggered Write & $-$Scatter Fusion & $-$All \\
        \midrule
        Time (\textmu s) & 37714.1 & 41125.7 & 48141.5 & 51053.5 \\
        Vector Ratio (\%) & 24.79 & 22.69 & 19.55 & 18.23 \\
        Scalar Ratio (\%) & 44.09 & 42.68 & 57.67 & 57.22 \\
        MTE2 (\%) & 24.06 & 21.97 & 15.79 & 14.31 \\
        MTE3 (\%) & 70.41 & 72.32 & 75.88 & 77.45 \\
        \bottomrule
    \end{tabular}
\end{table}

For the forward pass, disabling either Adaptive VecLen or Gather Fusion leads to a clear increase in kernel time and scalar unit utilization, indicating that both optimizations are crucial for maintaining high vectorization efficiency and reducing memory access overhead. When both are disabled, the performance degradation is more severe than the sum of individual effects, suggesting a strong synergy between these features. In training mode, the need for additional IO buffer space further reduces the effective vector length, making these optimizations even more important.

For the backward pass, disabling Staggered Write Scheduling or Scatter Fusion results in a significant increase in kernel time and memory transfer engine utilization, confirming their effectiveness in alleviating write contention and improving write bandwidth. The backward pass benefits most from these optimizations, achieving nearly $9\times$ speedup over the baseline when all are enabled. These results highlight the importance of co-designing algorithmic and hardware-aware optimizations for next-generation AI accelerators.

\section{Conclusion}

In this work, we present xMSDA, a high-performance implementation of MSDA on Ascend NPUs, enabled by a systematic system-algorithm co-design methodology. Our approach introduces a suite of hardware-aware optimizations, including the use of non-officially supported type-unaligned gather instructions, adaptive vectorization, padding-based alignment fixes, and contention-aware scheduling for scatter operations. Extensive experiments demonstrate that xMSDA achieves up to $5.9\times$ (forward), $8.9\times$ (backward), and $7.3\times$ (end-to-end training) speedup over the grid sample-based baseline, and up to $2.4\times$ over the latest vendor library. Our co-design paradigm and optimization strategies are broadly applicable to other attention operators and AI accelerators, providing a practical reference for future co-design efforts in the field.

\bibliographystyle{unsrt}
\bibliography{reference}

\end{document}